\documentclass[3p]{elsarticle}

\biboptions{sort&compress}

\journal{Arxiv}

\usepackage{amsmath}
\usepackage{xcolor}
\usepackage{mathtools}
\usepackage{soul}
\usepackage{setspace} %double spacing
\usepackage{float}

\usepackage{threeparttable}
\usepackage{etoolbox}
\usepackage{longtable}
\usepackage{multirow}
\usepackage{subfigure}    
\usepackage{booktabs}
\usepackage[utf8]{inputenc}
\usepackage{enumitem}
\usepackage{epsfig}
\usepackage{hyperref}

\usepackage{optidef}

\definecolor{darkgreen}{rgb}{0.1,0.8,0.1}
\newcommand{\fig}{Figure~}
\newcommand{\insitu}{\textit{in-situ }}

\newcommand{\degree}{^{\circ}}

\newcommand{\ra}[1]{\renewcommand{\arraystretch}{#1}}

\newcommand{\ts}{\textsuperscript}
\newcommand{\us}{\textsubscript}

\begin{document}
\singlespace
%\doublespace

\begin{frontmatter}

\title{Automated Identification of Slip System Activity Fields\\
from Digital Image Correlation Data}

\author[mymainaddress]{T. Vermeij}
\address[mymainaddress]{Dept. of Mechanical Engineering, Eindhoven University of Technology, 5600MB Eindhoven, The Netherlands}
\author[mymainaddress]{R.H.J. Peerlings}
\author[mymainaddress]{M.G.D. Geers}
\author[mymainaddress]{J.P.M. Hoefnagels*}
\cortext[mycorrespondingauthor]{Corresponding author}
\ead{j.p.m.hoefnagels@tue.nl}

\begin{abstract}
Crystallographic slip system identification methods are widely employed to characterize the fine scale deformation of metals. While powerful, they usually rely on the occurrence of discrete slip bands with clear slip traces and can struggle when complex mechanisms such as cross-slip, curved slip, diffuse slip and/or intersecting slip occur. This paper proposes a novel slip system identification framework, termed SSLIP (for Slip Systems based Local Identification of Plasticity), in which the measured displacement gradient fields (from Digital Image Correlation) are locally matched to the kinematics of one or multiple combined theoretical slip systems, based on the measured crystal orientations. To identify the amounts of slip that conforms to the measured kinematics, an optimization problem is solved for every datapoint individually, resulting in a slip activity field for every considered slip system. The identification framework is demonstrated and validated on an HCP virtual experiment, for discrete and diffuse slip, incorporating 24 slip systems. Experimental case studies on FCC and BCC metals show how full-field identification of discrete slip, diffuse slip and cross-slip becomes feasible, even when considering 48 slip systems for BCC. Moreover, the methodology is extended into a dedicated cross-slip identification method, which directly yields the orientation of the local slip plane trace orientation, purely based on the measured kinematics and on one or two chosen slip directions. For even more challenging cases revealing a persistent uncertainty in the slip identification, a two-step identification approach can be employed, as is demonstrated on a highly challenging HCP virtual experiment. DOI: \url{https://doi.org/10.1016/j.actamat.2022.118502}

\end{abstract}


\end{frontmatter}

\section*{Graphical Abstract}

\includegraphics[width=0.85\textwidth]{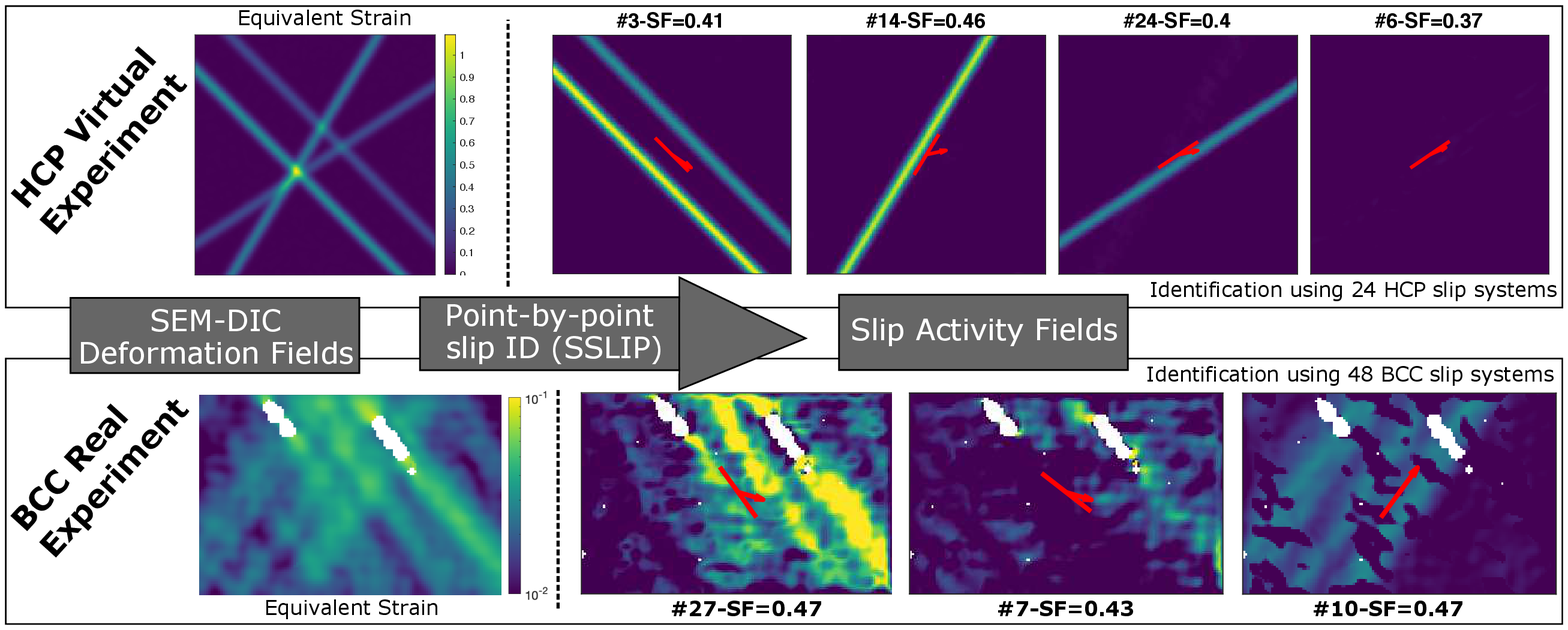}

\section{Introduction}
Plastic deformation in metals and engineering alloys occurs predominantly through the collective motion of dislocations, a phenomenon which is known as crystallographic slip. The directions in which slip may occur depend on the crystal structure and its orientation. They are generally described by a slip plane over which dislocations glide in the slip direction, whereby the combination of a slip plane and direction is referred to as a 'slip system'. Multiple slip directions, and hence slip systems, may be associated with a particular slip plane. Slip planes are often the most densely packed planes in the lattice, with the slip directions being the most closed-packed directions in each plane. Typical academic examples of the manifestation of crystallographic slip in experiments include discrete (sharp) and planar (straight) slip steps or lines, from a limited number of slip systems from one "symmetrically equivalent" slip family, e.g., 12 $\{111\}<\bar{1}10>$ systems in Face Centered Cubic (FCC) crystals, 12 $\{011\}<11\bar{1}>$ systems in Body Centered Cubic (BCC) crystals or 3 $\{0001\}<11\bar{2}0>$ systems in Hexagonal Close Packed (HCP) crystals. In reality, however, e.g. in engineering alloys or at higher temperatures, additional slip families may be activated, which can enhance or deteriorate the mechanical properties of interest \cite{tang2014formation, britton2015mechanistic, Tian2020}. Moreover, complex dislocation activities and limited spatial resolutions of microscopy techniques can introduce (apparent) mechanisms such as cross-slip, curved slip and diffuse slip \cite{lunt2018quantification, githens2020characterizing, harte2020effect, vermeij2022nanomechanical}.

The mechanical behaviour of crystalline materials in the plastic regime is inherently determined by the activation of and interaction between crystallographic slip systems. Therefore, the ability to identify active slip families, or even individual active slip systems, in experiments is of great interest for understanding, predicting and improving the (micro-)mechanical behaviour of materials. For a polycrystalline microstructure, (\textit{in-situ}) Scanning Electron Microscopy (SEM) can be employed to observe discrete and straight slip step traces. These can be matched to theoretical slip plane traces, based on an assumed-to-be-active slip family and the crystal orientation of the corresponding grain, measured through e.g. Electron Backscatter Diffraction (EBSD) \cite{bridier2005analysis, Bieler009, echlin2016incipient}. In the absence of clear slip steps, (equivalent) strain maps measured using SEM-based Digital Image Correlation (SEM-DIC) can also be employed for slip trace analysis \cite{OROZCOCABALLERO2017367, harr2021effect}.

However, to establish a complete picture of the activity of potential slip systems, identification of only the trace of the slip plane is rarely sufficient. There are usually multiple possible slip directions for the same slip plane trace, e.g., for different slip families (such as $<a>$ pyramidal and 1\ts{st} order $<c+a>$ pyramidal slip in HCP), for different slip planes with the same apparent slip plane trace, or even for different slip systems on the same slip plane. While the Schmid Factor (SF) can be employed to estimate the likelihood of activity of the respective slip directions \cite{bridier2005analysis}, this involves assumptions on the local stress state and knowledge of the Critical Resolved Shear Stress (CRSS) of a certain slip family, both of which are not trivially determined in experiments. Therefore, recent efforts have utilized DIC displacement data to retrieve the in-plane part of the slip direction for a certain slip step trace \cite{BOURDIN2018307,chen_daly_2016}. The "Heaviside DIC method" introduces an extra step function inside a DIC subset, whereby, for single, discrete and planar slip, the slip trace and in-plane slip direction are directly inferred from the DIC degrees of freedom associated with the step function parameters \cite{BOURDIN2018307}. Alternatively, the Relative Displacement Ratio (RDR) method can be employed on regular DIC data, whereby the ratio between the x- and y-component of the displacement, relative between both sides of a single and sharp slip trace, is compared to the ratio between the x- and y-component of the theoretical Burgers vectors that could be active for that specific slip trace \cite{chen_daly_2016}. Slip system identification through Heaviside DIC and RDR have been successfully employed in several works \cite{XU2019376, STINVILLE2020172, sperry2021comparison}, yet require the presence of sharp and discrete slip traces. As a result, in the event of diffuse slip, intersecting slip, cross-slip and/or curved slip, the true discrete slip events at the nanoscale are more often than not indiscernible, due to the limited spatial resolution of SEM-DIC, thereby restraining the applicability of the Heaviside DIC and RDR methods. This is unfortunate, because quantification of the activity of each slip system, in the form of a slip activity field for each possible slip system, could reveal local nanoscale variations of overlapping slip activity, e.g. near grain and phase boundaries, and can also facilitate direct comparisons between experiments and simulations. 

By using the rich deformation data that is produced through DIC, this paper presents a new automated method targeting point-by-point identification of all slip system activities. The proposed methodology, termed as SSLIP (for \textbf{S}lip \textbf{S}ystems based \textbf{I}dentification of \textbf{L}ocal \textbf{P}lasticity), operates by solving an optimization problem at every SEM-DIC datapoint in the map. The measured in-plane kinematics (in the form of displacement gradient fields) are matched to a combination of theoretical slip system kinematics (based on EBSD data), each with its own identified amplitude, yielding a slip activity field for each considered slip system.

In Section \ref{sec:concept}, we explore the feasibility of the SSLIP method and conceptualize the matching of measured and theoretical slip system kinematics. In Section \ref{sec:method}, the full SSLIP method is explained and validated using a virtual experiment on an HCP single crystal. Next, in Section \ref{sec:fcc}, experimental SEM-DIC data on an FCC Ni-based superalloy, taken from the literature \cite{harte2020effect, harte2020statistical, harte2020high}, is used to demonstrate the power of the local SSLIP method for discrete slip, including the ability to identify cross-slip. Subsequently, in Section \ref{sec:bcc}, an \insitu SEM-DIC micro-tensile test on a BCC ferrite specimen is analyzed, which presents a case study of diffuse slip and apparent cross-slip. This case study is highly challenging for the identification method since up to 48 slip systems need to be considered simultaneously. An extension of the SSLIP method on cross-slip identification is presented in Section \ref{sec:cross-slip}, in which we directly identify a local arbitrary slip plane trace orientation for a given slip direction. Finally, in Section \ref{sec:advanced}, we explore another virtual HCP case study which proves to be too challenging for the regular SSLIP method, but for which a successful identification is nevertheless achieved using a multi-step approach.

\section{Concept and Feasibility of SSLIP}
\label{sec:concept}

\begin{figure}[H]
    \includegraphics[width=1\textwidth]{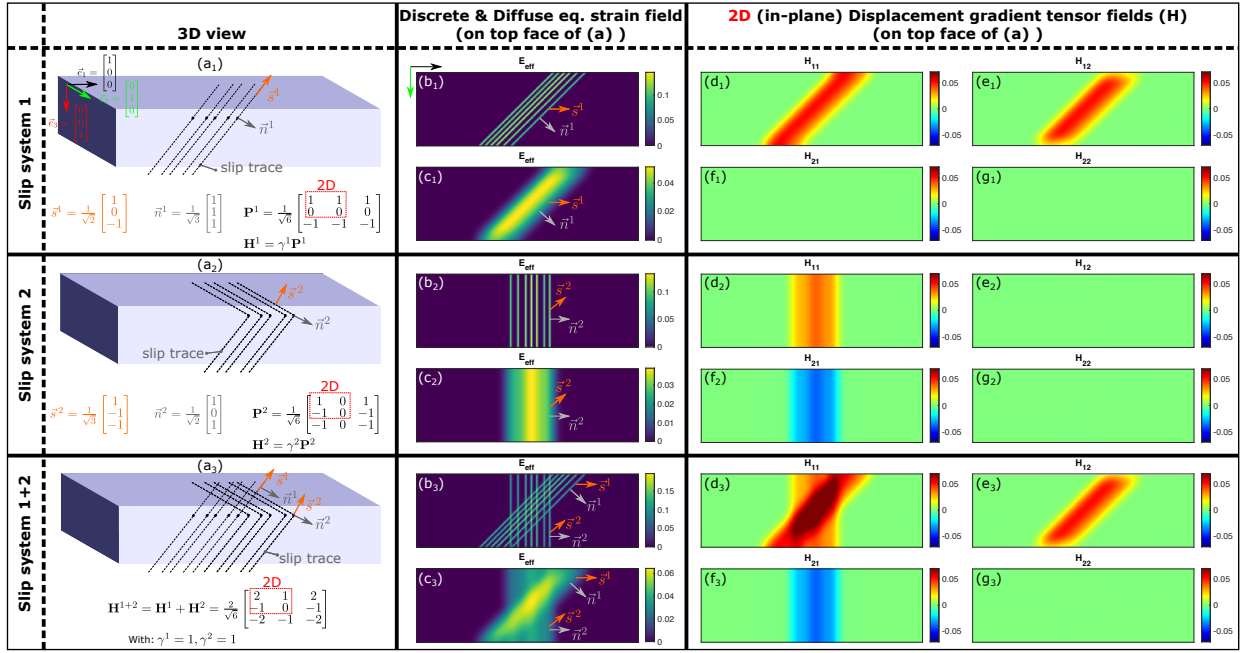}
    \caption{\small \textit{Schematic overview showing how diffuse deformation maps retain the information on the kinematics of, initially discrete, individual (systems 1 and 2 in (a) and (b) respectively) and overlapping (systems 1 and 2 combined in (c)) slip system activities. For, e.g., slip system 1 in (a): (a\us{1}) shows a crystal, in the undeformed configuration, with slip traces drawn on 2 sides through dashed lines, with slip direction $\vec{s\:}^1$ and slip plane normal $\vec{n}^1$ drawn in orange and grey respectively. (a\us{1}) also gives the 3D vector descriptions of $\vec{s\:}^1$ and $\vec{n}^1$, and the resulting theoretical displacement gradient tensor (Schmid tensor) $\mathbf{H}^1=\gamma^1\vec{s\:}^1 \otimes \vec{n}^1$ in matrix form. In (b\us{1}) and (c\us{1}), the frontside discrete and diffuse representations of the strain fields are given, followed by the 4 diffuse fields of the in-plane displacement gradient tensor components, $\mathbf{H}_{11}$, $\mathbf{H}_{12}$, $\mathbf{H}_{21}$ and $\mathbf{H}_{22}$ in (d\us{1}), (e\us{1}), (f\us{1}) and (g\us{1}) respectively. (b) and (c) follow the same structure for the subfigures.
    }}
    \label{fig1}
\end{figure}

We start by considering an idealized academic example of slip system activities to conceptually understand how the SSLIP method is feasible. This example will illustrate how the kinematics of single and intersecting slip bands are retained in diffuse deformation maps, forthwith allowing point-by-point identification of slip. \fig \ref{fig1}a\us{1},a\us{2} shows, for two distinct slip systems 1 and 2, multiple slip steps along a slip plane with normal $\vec{n}$, which leaves slip traces on the front and lateral surfaces, with a slip direction $\vec{s}$, which is the normalized Burgers vector. While the true kinematics at the atomic level always occur through discrete steps, continuum mechanics descriptions are employed, in, e.g., crystal plasticity frameworks, wherein the slip plane normal and slip direction are combined into the Schmid tensor $\mathbf{P}$, which, multiplied with a slip amplitude, contributes to the theoretical displacement gradient tensor $\mathbf{H}^{theor}$, as computed for slip systems 1 and 2, in \fig \ref{fig1}a\us{1} and \ref{fig1}a\us{2}, respectively. Next, we generate artificial displacement fields, for the top side of the crystals, that conform to the kinematics of slip systems 1 and 2. The "discrete" equivalent strain field, computed from the (slightly blurred) displacement fields, are shown in \fig \ref{fig1}b\us{1},b\us{2}, in which the individual slip bands can easily be discerned. However, limited spatial resolutions of microscopy techniques and filtering effects of local DIC smear out the discreteness, which is clarified by showing the equivalent strain fields in \fig \ref{fig1}c\us{1},c\us{2} for a diffuse deformation field, computed after major blurring of the displacement fields. 

While in most studies such a field of equivalent strain, or of a single strain component, is employed to identify slip bands and their trace orientation, the "measured" 2D (in-plane) displacement gradient tensor $\mathbf{H}^{exp}$ represents a more complete measure of the kinematics and can readily be computed by taking the gradient of the displacement fields. This is shown, for the diffuse case, as a field for each of the 4 in-plane components in \fig \ref{fig1}d\us{1},e\us{1},f\us{1},g\us{1} and \ref{fig1}d\us{2},e\us{2},f\us{2},g\us{2}, for systems 1 and 2 respectively. At locations of slip activity, these fields match the 2D (in-plane) part of the displacement gradient tensors computed in \fig \ref{fig1}a\us{1} and \ref{fig1}a\us{2}. More interestingly, we consider the case where slip systems 1 and 2 overlap in \fig \ref{fig1}a\us{3}, where we assume that, for the overlapping areas, the kinematics of slip systems 1 and 2 can be combined to describe the kinematics of the overlapping slip. While the discrete strain field (\fig \ref{fig1}b\us{3}) can still be used to distinguish the individual slip traces, the large area of overlapping slip in the diffuse strain field in \fig \ref{fig1}c\us{3} shows that methods such as the RDR and Heaviside DIC may not be applicable, since the slip traces cannot be identified. Yet, the in-plane displacement gradient tensor field in \fig \ref{fig1}d\us{3},e\us{3},f\us{3},g\us{3} reveal that the kinematics of slip systems 1 and 2 are still captured uniquely in components $H_{12}$ and $H_{21}$, while the $H_{11}$ component has a shared contribution. When multiple slip systems occur that are visually less distinguishable, or even provide a non-unique solution, we will show that their contribution to the kinematics can still be matched to the measured in-plane displacement gradient tensor field. Note that these kinematics are available in each point of the deformation data, thereby eliminating the need for visually discernible slip bands.

\section{Slip Systems based Local Identification of Plasticity (SSLIP) methodology}
\label{sec:method}

The SSLIP method will be elaborated and validated using a realistic virtual experiment on an HCP single crystal, for a discrete and diffuse case of 3 different, partially overlapping, active slip systems. Employing synthetic data generated in a virtual experiment allows us to assess the error made in the identification, since the exact solution is known in advance. 

The setup of the virtual experiment will be explained first (§\ref{sec:method1}). Next the method will be explained in three steps: the computation of the displacement gradient tensor from the DIC data (§\ref{sec:method2}), the derivation of the theoretical displacement gradient tensor (§\ref{sec:method3}), and the one-step point-by-point SSLIP optimization procedure (§\ref{sec:method4}). Finally, SSLIP is demonstrated on a discrete and diffuse version of the virtual experiment (§\ref{sec:method5},§\ref{sec:method6}). In the rest of the paper, the SSLIP method is also applied to real experimental measurement data in Sections \ref{sec:fcc}, \ref{sec:bcc} and \ref{sec:cross-slip}. 

\subsection{Virtual HCP experiment}
\label{sec:method1}

The virtual SEM-DIC data is generated on a square domain with a single crystal orientation. For HCP materials such as Zn and Mg, Basal slip dominates under most loading conditions \cite{britton2015mechanistic}. However, under highly constrained conditions (e.g. near a grain boundary) or at higher temperatures, alternative slip families can be activated in the form of 1\ts{st} and 2\ts{nd} order Pyramidal $<c+a>$ (cross-)slip \cite{tang2014formation, bell1957dynamics}. Therefore, for a single crystal orientation, we choose to activate the slip system with the highest SF (for horizontal tension) from each of these 3 slip families, in the form of slip bands. A displacement field is generated on a grid of 200 by 200 pixels by combining 4 individual slip steps: two occurrences of the same Basal slip system ($\{0001\}<11\bar{2}0>$; $\#3_a$ \& $\#3_b$) with different amplitudes, one 1\ts{st} order $<c+a>$ Pyramidal slip system ($\{10\bar{1}1\}<\bar{1}\bar{1}23>$; $\#14$) and one 2\ts{nd} order $<c+a>$ Pyramidal slip system ($\{11\bar{2}2\}<\bar{1}\bar{1}23>$; $\#24$), with both pyramidal systems having the exact same slip direction (for additional complexity). The displacement field for each single slip event is generated by defining its slip trace, at a certain position, under the correct angle, and assigning displacement values according to the in-plane Burgers vector, multiplied by a certain slip amplitude, on the right side of the slip trace. The 4 single slip displacement fields are then superposed to form a combined displacement field. To simulate experimental conditions, Gaussian noise with a standard deviation of $0.05$ pixel is added (which is relatively large considering that the noise in the DIC displacement data is often as low as $0.01$ pixel) and subsequent Gaussian filtering is applied to emulate SEM-DIC conditions under which we recently showed to achieve high spatial strain resolutions \cite{vermeij2022nanomechanical, Vermeij2021}. See Table \ref{Table:virtExp1} for details on the generation of this virtual experiment.
The x-component of the resulting virtually generated displacement field $\vec{u}$ is shown in \fig \ref{fig2}a.

\begin{table}[H]\centering
\caption{\textit{\small All relevant details for the generation of the HCP virtual experiment.}}
\label{Table:virtExp1}
\ra{1.3}
\begin{tabular}{@{}| l | c r|@{}}\toprule
%\begin{tabular}
& \textbf{Value} & \textbf{Unit}\\
\midrule
Gaussian Noise Standard Deviation (STD)        & $0.05$ & pixel\\
Gaussian filter STD        & $3$ & pixel\\
Crystal Orientation Euler Angles (ZXZ)        & $[45,45,15]$ & $\degree$\\
Amplitude $\#3_a$; Basal $(0001)[2\bar{1}\bar{1}0]$ & $\frac{2}{3}$ & pixel\\
Amplitude $\#3_b$; Basal $(0001)[2\bar{1}\bar{1}0]$ & $\frac{4}{3}$ & pixel\\
Amplitude $\#14$; 1\ts{st} order $<c+a>$ Pyramidal $(\bar{1}101)[\bar{2}11\bar{3}]$ & $1$ & pixel\\
Amplitude $\#24$; 2\ts{nd} order $<c+a>$ Pyramidal $(\bar{2}112)[\bar{2}11\bar{3}]$ & $\frac{2}{3}$ & pixel\\
\bottomrule
\end{tabular}
\end{table}

\subsection{Computation of the displacement gradient tensor field}
\label{sec:method2}
\begin{figure}[H]
%	\centering
    \includegraphics[width=0.7\textwidth]{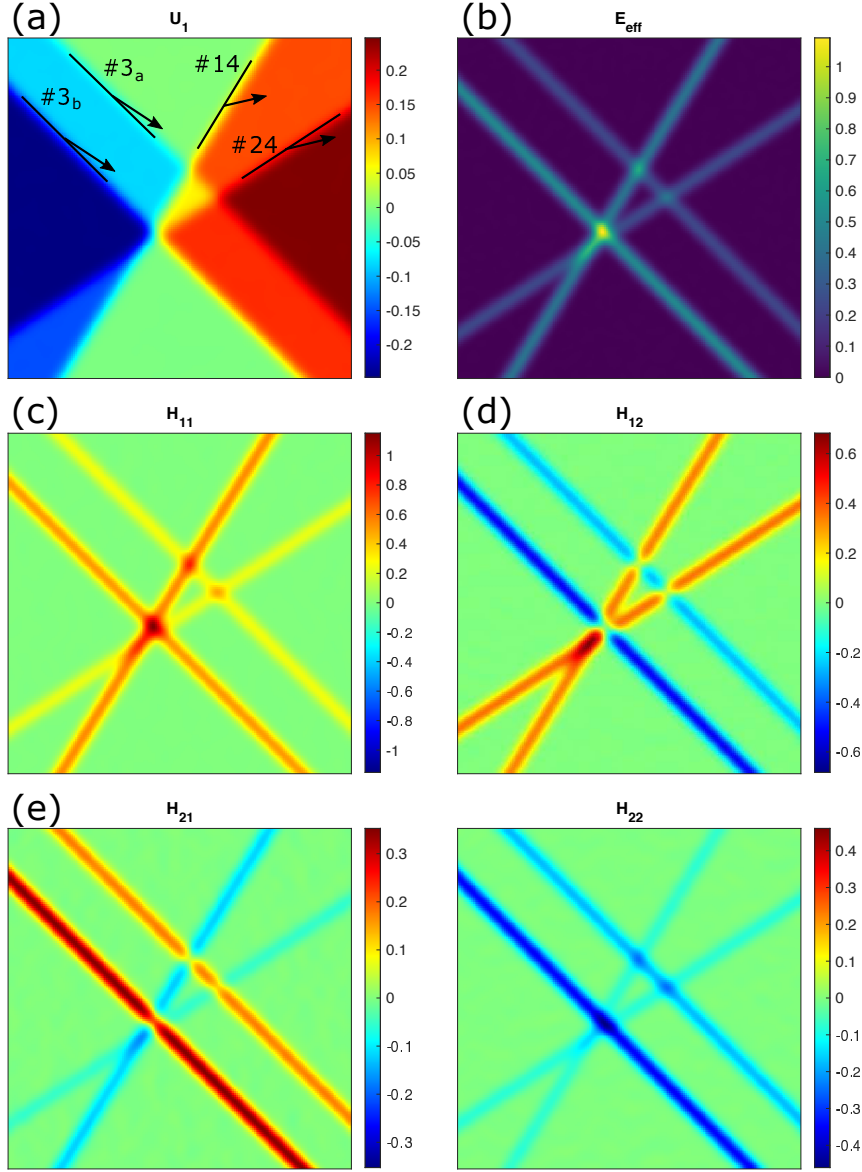}
    \caption{\small \textit{Virtual HCP SEM-DIC experiment, with (a) the generated displacement field (x component) with slip trace and direction indicated and slip system numbers given, and (b) the effective strain field $E_{eff}$. The 4 in-plane displacement gradient tensor components $H_{11}$, $H_{12}$, $H_{21}$ and $H_{22}$ are plotted in (c), (d), (e) and (f) respectively.
    }}
    \label{fig2}
\end{figure}

The displacement data needs to be converted to in-plane deformation data that can be compared to slip system kinematics. Therefore, we compute the displacement gradient tensor $\mathbf{H}^{exp}$ by taking the in-plane gradient of $\vec{u}$, through central differences:
\begin{equation}
\mathbf{H}^{exp}=\vec{\nabla}_0\vec{u},
\end{equation}
from which the effective strain $E_{eff}$, which is an equivalent, shear-dominated, strain measure that reflects the occurrence of slip, can be computed \cite{dieter1976mechanical}:
\begin{equation}
E_{eff} = \sqrt{\frac{1}{2} (H_{11}-H_{22})^2 + \frac{1}{2} (H_{12}+H_{21})^2},
\end{equation}
which is shown in \fig \ref{fig2}b. While this effective strain map is suitable to visually identify the slip bands and their traces, the maps of the 4 components of the displacement gradient tensor $\mathbf{H}^{exp}$ in \fig \ref{fig2}c-f reveal, similarly to \fig\ref{fig1} in the Introduction, how the local kinematics clearly differ between the different slip systems, especially for the shear terms $H_{12}$ and $H_{21}$. Note how the pyramidal slip bands (systems $\#14$ and $\#24$), which deform along the same slip direction but over a different slip plane, differ in their $H_{11}$ and $H_{21}$ components, yet have equal $H_{12}$ and $H_{22}$ components, showing how the difference in slip trace is retained at every datapoint in the map.

\subsection{From theoretical slip systems to (in-plane) deformation kinematics}
\label{sec:method3}
In order to use the measured displacement gradient tensor components for slip system identification, we require a similar description for (a combination of) theoretical slip systems. For a single slip system $\alpha$, its contribution to the displacement gradient tensor, $\mathbf{H}^{\alpha}$ in the undeformed configuration, can be written as the slip amplitude $\gamma^{\alpha}$ multiplied with the Schmid tensor $\mathbf{P}^{\alpha}$:
\begin{equation}
\mathbf{H}^{\alpha}=\gamma^{\alpha} \mathbf{P}^{\alpha}=\gamma^{\alpha}\vec{s}^{\:\alpha} \otimes \vec{n}^{\alpha},
\end{equation}
in which the dyadic product is taken between the vectors $\vec{s}^{\:\alpha}$ and $\vec{n}^{\alpha}$, which are the normalized Burgers vector and slip plane normal, respectively, of slip system $\alpha$. In the case of multiple active slip systems, we find the total theoretical displacement gradient tensor by addition over each slip system:
\begin{equation}
\mathbf{H}^{theor}=\sum_{\alpha=1}^{N} \mathbf{H}^{\alpha},
\end{equation}
in which $N$ is the total number of slip systems. This formulation is similar to how the plastic velocity gradient tensor is formulated in crystal plasticity (CP) simulations. However, we do employ two assumptions here:
\begin{itemize}
\item All deformation is plastic only. Thereby, elastic strains are neglected, which seems justified since elastic strains usually fall well below the noise level of SEM-DIC and far below the plastic strain levels.
\item We choose to formulate the framework in a small deformations configuration. This is required because the resulting simplification of the combination of kinematics from different slip systems critically lowers the computational costs of the optimization problem, allowing it to be solved at every datapoint separately.
\end{itemize}

\subsection{One-step point-by-point SSLIP method}
\label{sec:method4}
The core of the identification method lies in the assumption that the in-plane displacement gradient tensor components, computed directly from the displacement fields, corresponds to the in-plane deformation kinematics of a combination of active slip systems. 

For every data point in the displacement gradient tensor map, we solve an optimization problem to find the combination of slip amplitudes $\gamma^{\alpha}$ that best matches the measured displacement gradient tensor. While one could argue that the objective function should be the displacement gradient tensor residual L\us{2} norm (of the in-plane part of $\mathbf{H}$, written as $||\hdots||^{2D}$), i.e. $||\mathbf{H}^{exp}-\mathbf{H}^{theor}||^{2D}$, minimization of this scalar value results in an underdetermined set of linear equations for full slip system identification, even when "only" 12 possible slip systems (e.g. for FCC, BCC or HCP) are included, since there are only 4 measured (known) components of the in-plane 2D displacement gradient tensor. Instead, we construct the following constrained optimization problem in which the total amplitude of the slip over all slip systems is minimized:
\begin{mini!}|1|
{\gamma^{\alpha}=\gamma^{1},\ldots,\gamma^{N}}{\sum_{\alpha=1}^{N}{|\gamma^{\alpha}|}\label{eq:objfun}}
{\label{eq:optim}}{}
\addConstraint{||\mathbf{H}^{exp}-\mathbf{H}^{theor}||^{2D}}{<H_{thresh}\label{eq:con1}}{}
%\addConstraint{\gamma^{\alpha}}{\geq0,\quad}{\alpha=1,\ldots,N,\label{eq:con2}}
\end{mini!}
which is motivated by the following considerations:
\begin{itemize}
\item The sum of 3D slip amplitudes is minimized to ensure that the measured 2D kinematics is described by the smallest amount of plastic slip, ensuring that two "in-plane identical" slip systems can still be discerned. An alternative choice could be to minimize the plastically dissipated energy, which would involve simply multiplying each slip amplitude with the CRSS of that slip system, in Eq. \ref{eq:objfun}. This would only be relevant when multiple slip families are considered and would be similar to the formulation of the plastically dissipated energy in CP simulations. However, such a formulation would ignore the strong dependence of the initiation of local slip on the presence of dislocation sources in polycrystalline materials, and could thereby favor slip systems that cannot activate when a dislocation source is not present. Therefore, we believe that the current minimization of total 3D plastic slip yields a more objective matching of the experimentally observed 2D kinematics. (Eq. \ref{eq:objfun})

\item We minimize the sum of the absolute values of the slip amplitudes in order to allow "backward" and "forward" slip and to make the SSLIP framework independent of the definition of the slip system, i.e. the sign of the slip plane normal and/or slip direction. While the "positive" or "forward" slip direction can easily be determined under simple loading conditions such as uniaxial tension, complex local or global boundary conditions could result in uncertainties in these directions, a problem that is circumvented by using the absolute values. (Eq. \ref{eq:objfun})

\item The residual L\us{2} norm, only computed for the in-plane part of the displacement gradient tensors, $||\mathbf{H}^{exp}-\mathbf{H}^{theor}||^{2D}$, is not required to vanish but is constrained to a value below a certain threshold, $H_{thresh}$, corresponding to 2 times the standard deviation of the statistical variation in $\mathbf{H}$ as determined from experiments, which depends on the SEM-DIC noise as well as systematic uncertainties in, e.g., crystal orientations and alignment. By analyzing the statistical variation of $\mathbf{H}^{exp}$ from a series of undeformed images, the $H_{thresh}$ can be determined. It should be noted that the identification is not very sensitive to the threshold value as long as the plastic strains are above the SEM-DIC noise level. (Eq. \ref{eq:con1})

\end{itemize}
This optimization problem can be solved numerically for every data point. In this work, we have used $Coneprog$ in $Matlab$ \cite{MATLAB:2010}, while (the slower) $Fmincon$ yields equal results. Additionally, all crystallographic and slip system data manipulation and all plotting has been performed using the MTEX toolbox \cite{MTEX, bachmann2010}. The full $Matlab$ code for the SSLIP method is available at: \url{https://www.github.com/TijmenVermeij/SSLIP}.

\subsection{Slip system identification on a discrete HCP virtual experiment}
\label{sec:method5}

\begin{figure}[H]
%	\centering
    \includegraphics[width=1\textwidth]{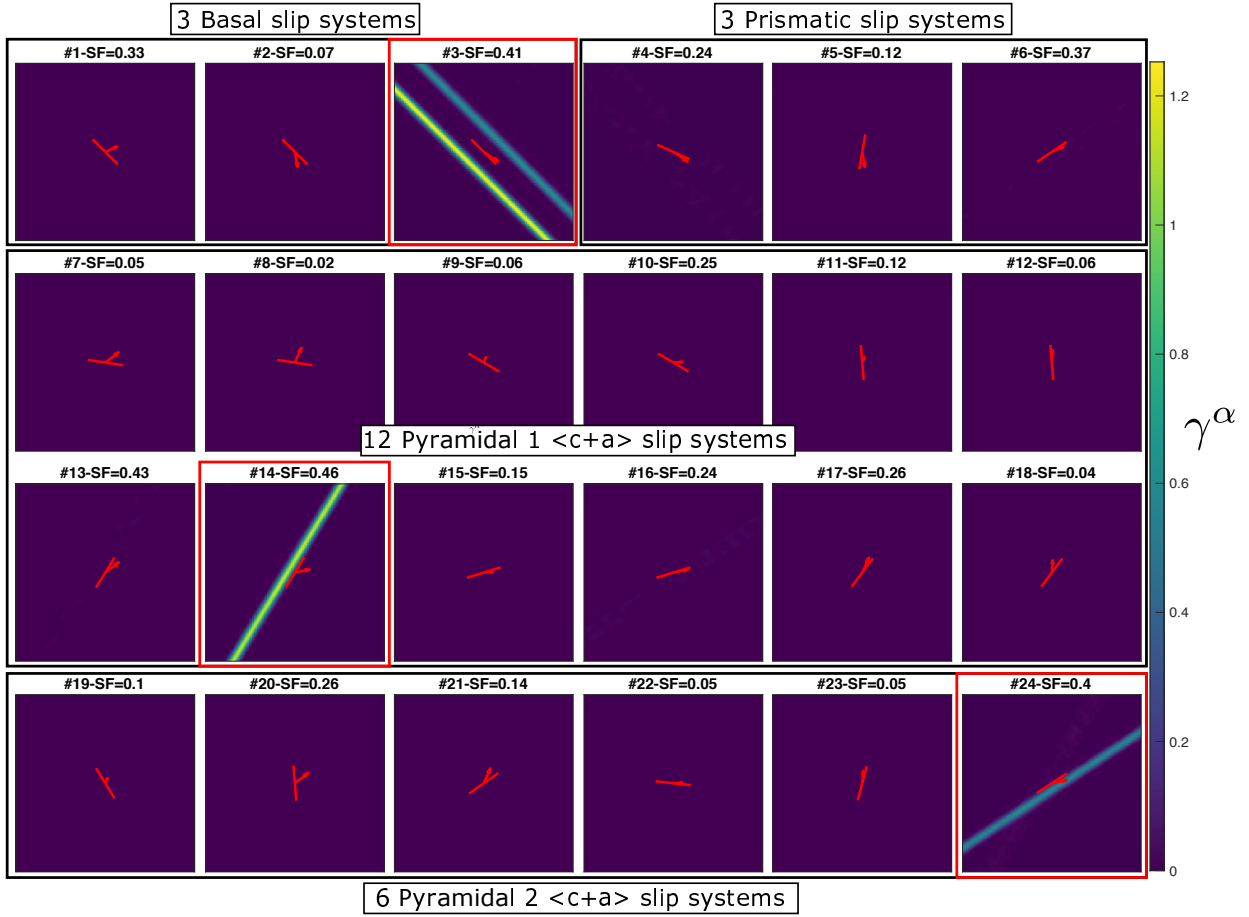}
    \caption{\small \textit{Slip system activity maps resulting from the point-by-point slip system identification for the HCP virtual experiment. For every possible slip system $\alpha$, grouped into 3 Basal, 3 Prismatic, 12 1\ts{st} and 6 2\ts{nd} order $<c+a>$ Pyramidal systems, a field of the identified slip amplitude $\gamma^\alpha$ is plotted. Slip systems $\#3$, $\#14$ and $\#24$ were the active systems in the virtual experiments and are highlighted by a red rectangle. For every slip system activity field the theoretical slip trace is drawn as a red line and the in-plane slip direction as a red arrow. }}
    \label{fig3}
\end{figure}

For the introduced virtual experiment, the SSLIP is performed without using any prior knowledge on the active slip systems. Therefore, the following 24 possible slip systems are included: 3 Basal, 3 Prismatic, 12 1\ts{st} and 6 2\ts{nd} order $<c+a>$ Pyramidal slip systems. The identified slip amplitude field $\gamma^\alpha$, for every slip system, is shown in \fig \ref{fig3}, with the theoretical slip trace and in-plane slip direction drawn in red. The slip systems that were triggered in the virtual experiment, $\#3$, $\#14$ and $\#24$, are accurately captured in the identification, as shown with the red rectangles around the slip amplitude fields. It is important to note that this is the result of a point-by-point identification which does not use any geometrical information of the slip band pattern (such as a trace orientation). Accordingly, even the overlapping parts of the slip bands are correctly identified, whereas none of the passive slip systems show any activity in the identification.

\subsection{Slip system identification on a diffuse HCP virtual experiment}
\label{sec:method6}
\begin{figure}[H]
%	\centering
    \includegraphics[width=0.7\textwidth]{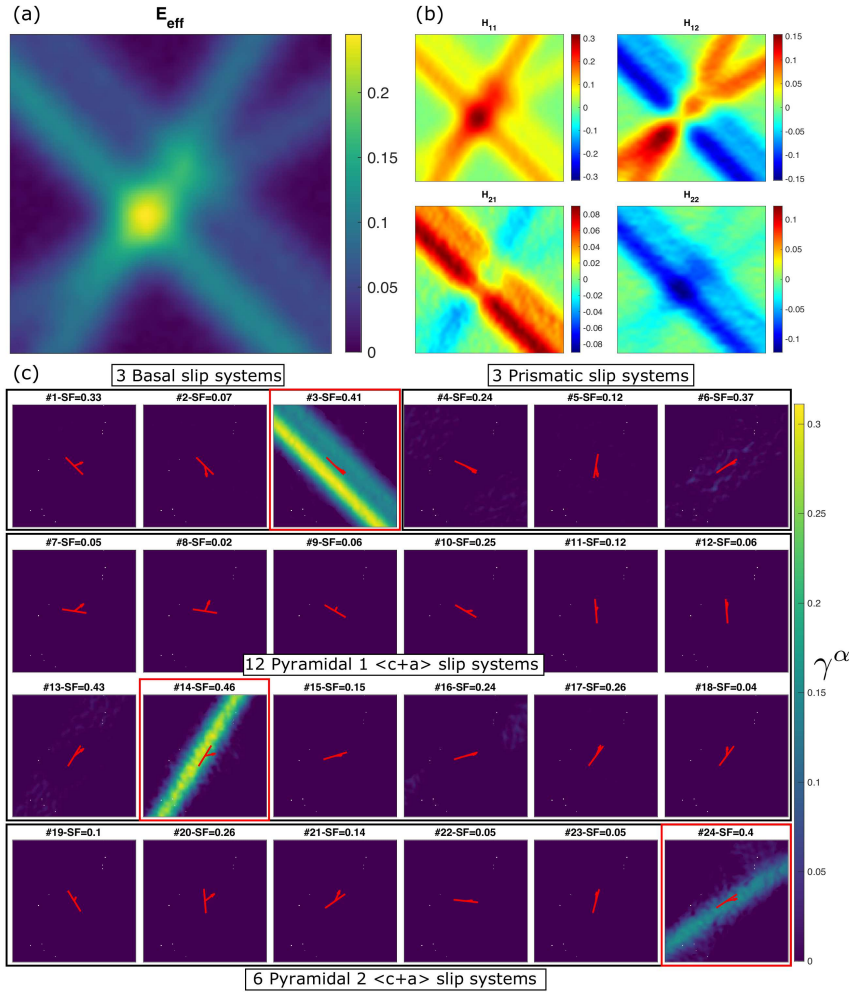}
    \caption{\small \textit{ A \textit{diffuse} HCP virtual experiment and the point-by-point slip system identification results. (a) Effective strain field and (b) in-plane displacement gradient tensor components. (c) Identified slip amplitude fields, with the applied slip systems $\#3$, $\#14$ and $\#24$ highlighted by red rectangles.  }}
    \label{fig4}
\end{figure}

Since it would be highly beneficial if the slip identification method is suitable for diffuse slip, we test this by applying a significant amount of blurring (Gaussian standard deviation of 12 pixels) on the displacement fields (before applying the Gaussian noise with subsequent DIC filtering to emulate SEM-DIC conditions as mentioned above), resulting in an effective strain map and displacement gradient tensor fields as shown in \fig \ref{fig4}a,b. Here, we demonstrate that the SSLIP framework does not require identification of geometric features (such as the trace orientation of a sharp slip band) from (equivalent) strain maps. Additionally, in this virtual experiment, there is considerably more overlap between the slip bands, however, the displacement gradient tensor field still reveals clear differences. Performing the identification on this diffuse virtual experiment recovers again the correct slip activity fields, see \fig \ref{fig4}c. The large deviations in displacement gradient intensity at the location of the overlap of the diffuse slip bands could suggest that a proper identification at overlapping locations is challenging. Nevertheless, the results show that the proposed SSLIP method has no difficulty distilling the right amounts of slip also in these regions and hence is quite robust to wide, overlapping slip bands. This gives confidence in the method for complex real-world experiments where multiple slip systems may overlap in a diffuse manner.

\section{FCC Ni-based superalloy case study: planar, cross and diffuse slip}
\label{sec:fcc}
\begin{figure}[H]
%	\centering
    \includegraphics[width=0.85\textwidth]{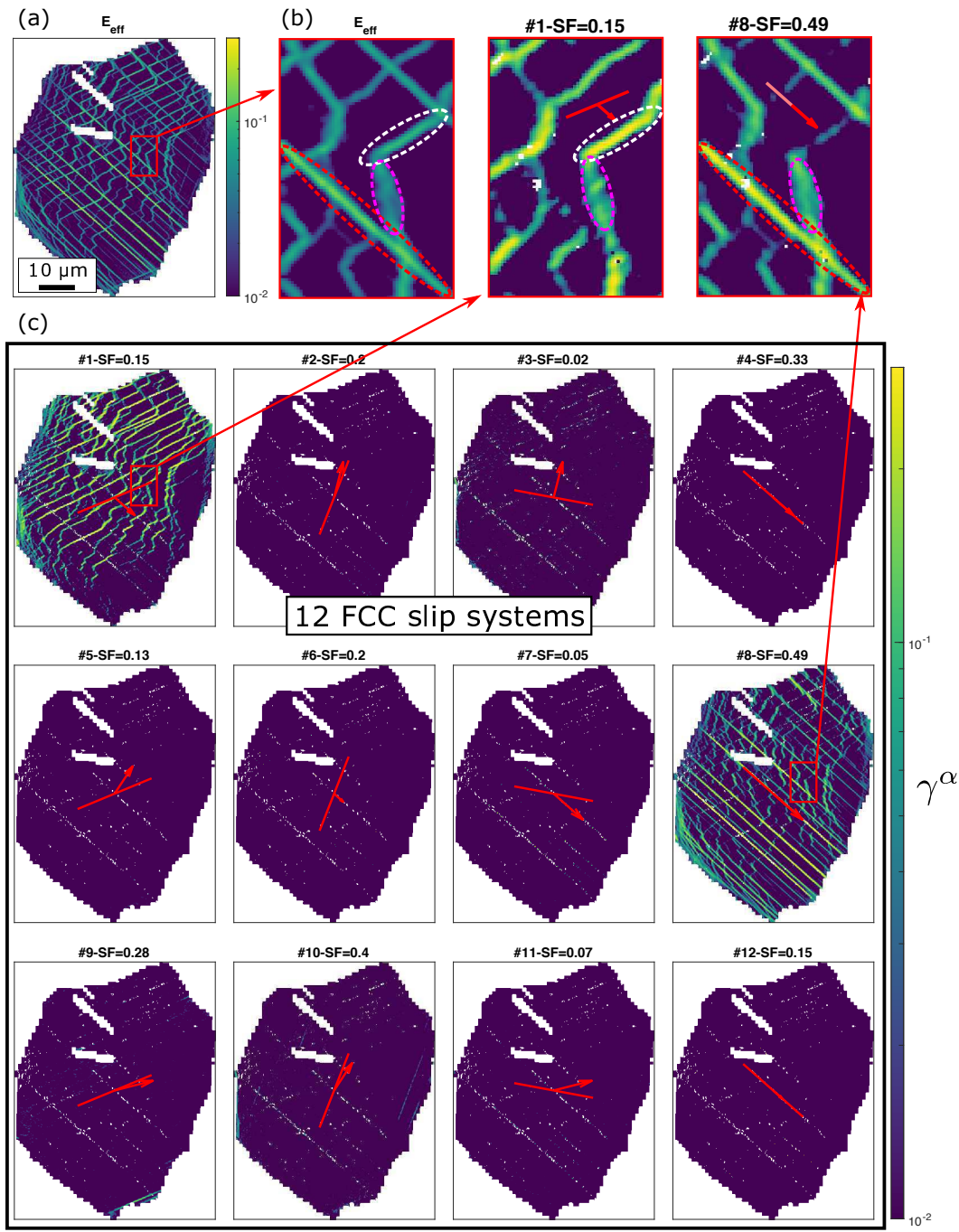}
    \caption{\small \textit{Slip system identification on grain 1 of a FCC Ni-based super alloy. (a) Effective strain field, (b) inset of effective strain field and slip activities of slip system $\#1$ and $\#8$ in a zoomed region, wherein the pink ellipse shows a slip band that is active in both slip systems (likely due to cross-slip), while the red and white ellipse indicate slip bands only present in the activity maps of systems $\#1$ and $\#8$, respectively. (c) Slip activity fields of all the 12 slip systems used in the identification.
    }}
    \label{fig5}
\end{figure}

\begin{figure}[H]
%	\centering
    \includegraphics[width=1\textwidth]{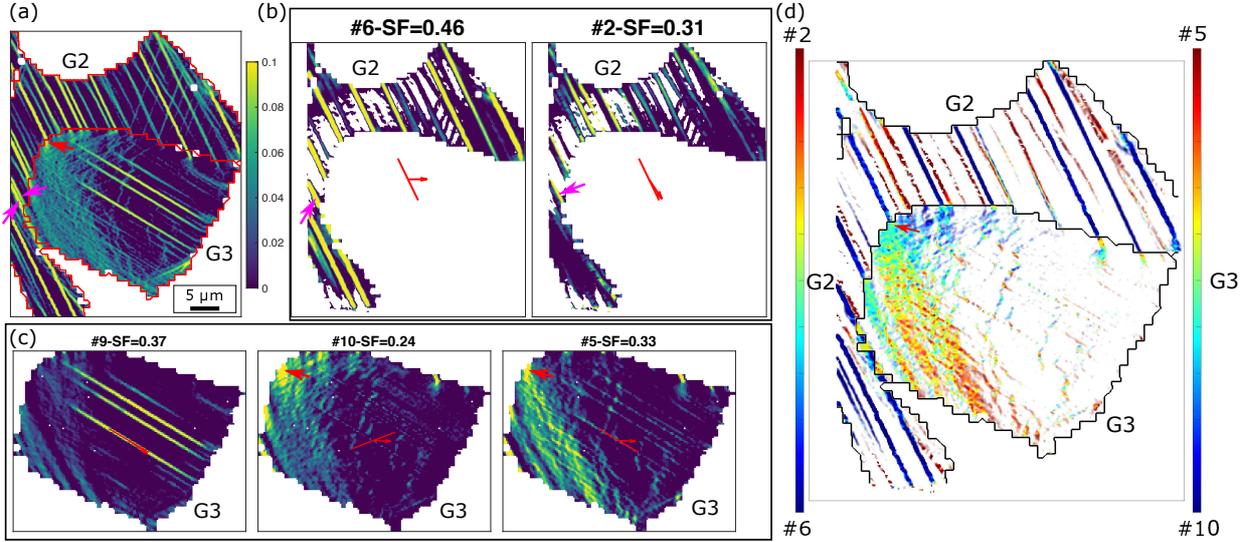}
    \caption{\small \textit{ Slip system identification on neighbouring Grains 2 and 3 of FCC Ni-based superalloy. (a) Effective strain field with grain boundaries overlaid in red and (b) most active identified slip systems of Grain 2 (systems $\#2$ and $\#6$), with pink arrows indicating 2 slip bands that are identified to have different slip directions. (c) The 3 most active identified slip system activity fields (systems $\#9$, $\#10$ and $\#5$) of Grain 3. In (d), the fraction of two identified slip system amplitudes is plotted for each Grain ($\#2$ and $\#6$ for Grain 2 and $\#5$ and $\#10$ for Grain 3). The transparency is scaled (inversely) according to the total slip amplitude. The red arrows in (a,c,d) indicate overlapping activity of systems $\#2$ and $\#6$ near the grain boundary.
    }}
    \label{fig6}
\end{figure}

For the experimental validation we start with a comparatively simple case study of mostly discrete slip lines in FCC, in which the number of possible slip systems is limited to 12 systems from the same slip family. Note that, while we perform a point-by-point identification, we can use the slip trace orientations here to verify that the correct slip systems have been identified, which can be valuable when possible, but is not a requirement.  A large dataset of SEM-DIC displacement maps and EBSD orientation maps, on a Ni-based superalloy RR1000, deformed under uniaxial tension to a global strain of $\sim0.02$, acquired by Harte \textit{et al.} \cite{harte2020effect, harte2020statistical}, was taken from an open-source data repository \cite{harte2020high}. For a polycrystalline or multi-phase microstructure, proper alignment is required between the EBSD microstructure and SEM-DIC displacement field data to ensure the crystal orientation and its corresponding slip system kinematics are known for each grain. Here, we employ our own data alignment framework \cite{vermeij2022nanomechanical} to align these fields. We selected two grains that show more complex deformation (i.e. Grain 1 shown in \fig \ref{fig5} and neighbouring Grains 2 and 3 in \fig \ref{fig6}) as interesting examples to demonstrate the efficacy of the new slip identification framework. For both areas, a full slip system identification using SSLIP is performed, considering all 12 $\{111\}<\bar{1}10>$ FCC slip systems. In this experiment, strains reach up to $0.3$ and therefore also serves well to test our small deformations assumption.

\fig \ref{fig5}a shows the effective strain field of Grain 1, in which discrete slip lines can be identified predominantly over 2 distinct traces. However, upon close inspection it is clear that these 2 slip traces are also connected at various locations, indicating that cross-slip may be active here. After identification, the 12 slip amplitude fields are plotted in \fig \ref{fig5}c, showing 2 dominant slip systems: $\#1$ and $\#8$. The activity of slip system $\#8$ could have been expected given its high SF ($0.49$), yet, slip system $\#1$ is not the slip system with the highest SF with that particular slip trace orientation. Slip systems $\#1$ and $\#8$ do, however, have the same slip direction, indicating that cross-slip could be active. Indeed, when comparing the theoretical slip traces of $\#1$ and $\#8$ overall, they match rather well to the apparent slip traces in the activity maps of $\#1$ and $\#8$, but not everywhere. Consider the inset in \fig \ref{fig5}b, showing the effective strain field and activity fields of slip systems $\#1$ and $\#8$ in a zoomed region. The connecting slip band, as indicated by the pink ellipse, appears to have an approximately equal amplitude in the slip system activity fields of $\#1$ and $\#8$, as well as a slip trace direction that is in between the theoretical slip trace direction of $\#1$ and $\#8$. In contrast, other slip lines that do follow the theoretical slip trace direction of system $\#1$ and $\#8$, indicated by the white and red ellipse respectively, are only present in one of the two amplitude maps. Therefore, identification of cross-slip becomes feasible using the proposed slip system identification method.

\fig \ref{fig6} shows a case of even more complex behaviour wherein interactions across a grain boundary appear to introduce diffuse slip. In \fig \ref{fig6}a, the effective strain maps for Grains 2 and 3 are plotted, overlaid with the grain boundary in red. Grain 2 predominantly shows discrete slip, which appears to be transferred into Grain 3 in a diffuse manner. A separate slip system identification is performed on both grains and the strongest slip system activity fields are shown in \fig \ref{fig6}b for Grain 2 and \fig \ref{fig6}c for Grain 3. First note that the white areas in \fig \ref{fig6}b show pixels for which the required threshold value $H_{thresh}$ in the optimization algorithm of the SSLIP method, see Eq. \ref{eq:optim}, has not been reached. Since the identification is not trustworthy for these data points, they have been omitted from the maps. Whereas Grain 2 mainly reveals slip bands along the same slip trace, the identification clarifies that there are actually 2 different slip systems active, with the same slip trace but different slip direction (i.e. systems $\#2$ and $\#6$). Even for slip bands that are very close to each-other, as indicated by the pink arrows in \fig \ref{fig6}a-b, the SSLIP method adequately separates these, which is expected to be a challenge for more traditional methods where the slip bands need to be clearly separated from each-other. Grain 3 (\fig \ref{fig6}c) shows a more complex behaviour. Slip system $\#9$ has the highest SF and forms clear slip bands, away from Grain 2. Slip systems $\#5$ and $\#10$ have the same slip direction (different from slip system $\#9$) and both slip systems activate in a diffuse manner, in approximately the same area. Again, this is a strong indication of cross-slip, yet is less obvious as compared to Grain 1, purely based on the strain maps, since there are no clearly visible slip bands along the theoretical slip traces. The occurrence of this dual-slip mechanism, which is likely cross-slip, appears to be especially prevalent at the grain boundary, where a slip band of Grain 1 impinges, as indicated by the red arrows in \fig \ref{fig6}a,c. Note that, near grain boundaries, elastic strains may also be considerable (while assumed to be negligible, see Section \ref{sec:method3}), e.g. in the absence of dislocation sources, which could affect the identification here. Therefore, we recommend that, for these complex diffuse cases, the results are carefully checked for inconsistencies with respect to the total slip activity in the grain. Next, in order to quantify the local relative contributions of 2 slip systems, we plot the fraction of the two relevant slip systems in a colormap in \fig \ref{fig6}d, both for Grains 2 and 3. The transparency is scaled inversely to the total slip amplitude. For Grain 2, most slip activity belongs completely to one of the two systems (i.e. the color is either red or blue), while for Grain 3 the ratio ranges over the full color scale, with a clear 'mixture' of slip systems $\#5$ and $\#10$ occurring where system $\#6$ of Grain 2 impinges the grain boundary (red arrow in \fig \ref{fig6}d). These observations illustrate the depth of analysis enabled by the proposed identification method.

\section{BCC ferrite case study: diffuse \& cross slip}
\label{sec:bcc}
\begin{figure}[H]
%	\centering
    \includegraphics[width=1\textwidth]{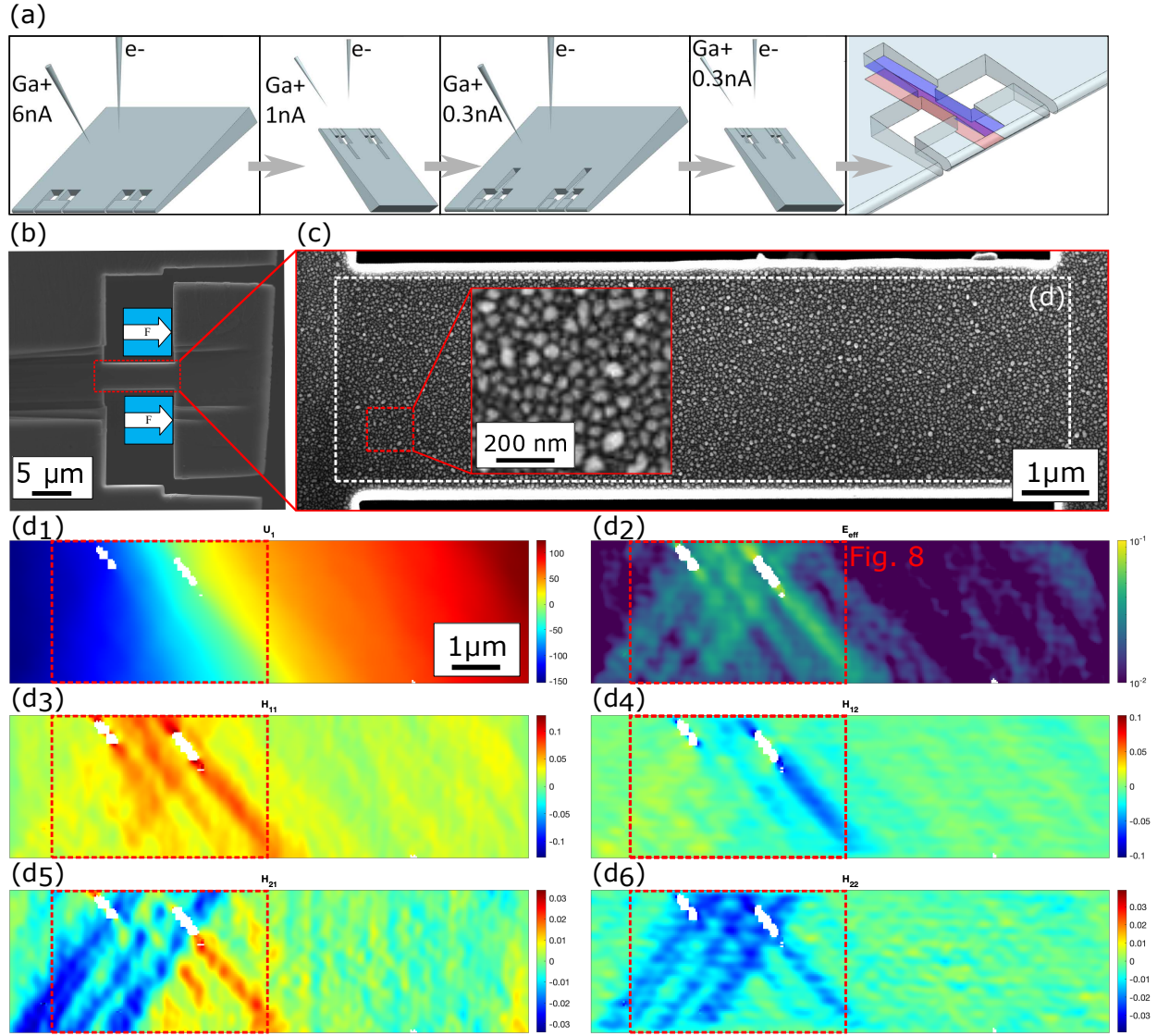}
    \caption{\small \textit{Overview of the \insitu SEM-DIC micro-tensile test on a single crystal BCC Ferrite specimen. (a) Schematic overview of the FIB milling procedure, (b) specimen overview with loading direction \cite{DuTensileTest} and (c) specimen after InSn nanoscale DIC patterning \cite{HoefnagelsPattern}, with an inset showing the pattern in detail. The white dashed rectangle indicates the region of interest used for at the final deformation increment, for: (d\us{1}) the x-component of the displacement field, (d\us{2}) the effective strain field and (d\us{3}-(d\us{6})) the in-plane displacement gradient tensor components $H_{11}$, $H_{12}$, $H_{21}$ and $H_{22}$. While slip system identification has been performed over the full specimen, the red dashed rectangle indicates the zoom area for which the activity fields will be shown in \fig\ref{fig8}. In the white areas, the DIC correlation was not successful.
    }}
    \label{fig7}
\end{figure}

Next, we test the SSLIP method on a challenging case study of diffuse slip in Ferrite (BCC) in which 48 slip systems are equally likely to activate, namely 12 $\{011\}<11\bar{1}>$, 12 $\{112\}<11\bar{1}>$ and 24 $\{123\}<11\bar{1}>$ systems \cite{Tian2020}. In order to establish a well-defined loading state, an \insitu SEM-DIC micro-tensile test was performed on a single crystal Ferrite specimen (extracted from a Dual-Phase steel microstructure), according to the methodology described in Ref. \cite{vermeij2022nanomechanical}. See \fig \ref{fig7}a for a schematic drawing of the Focused Ion Beam (FIB) milling procedure used to extract a tensile specimen from a ultra-thin wedge tip \cite{DuTensileTest}, with the resulting tensile specimen and the loading direction shown in \fig \ref{fig7}b. \fig \ref{fig7}c shows the specimen after application of an InSn nanoscale DIC speckle pattern, which is especially well-suited for these SEM-DIC micro-tensile tests \cite{HoefnagelsPattern, vermeij2022nanomechanical}.

The SEM-DIC data for the final deformation step is shown in \fig \ref{fig7}d in the form of the \textit{x}-component of the displacement field, the effective strain field and the 4 components of the in-plane displacement gradient tensor field. While the effective strain field does give indication of the presence of 2 different slip traces, there is significant overlap and the slip bands are rather diffuse. However, the displacement gradient tensor field, especially the $H_{21}$ component, clearly shows different kinematics for the different slip bands, similarly to the schematic illustration of the method in \fig\ref{fig1}. We therefore expect there to be at least 2 significantly different active slip systems. To identify all active slip systems, we run the SSLIP method, including all 48 slip systems simultaneously, for the full specimen area.

\begin{figure}[H]
%	\centering
    \includegraphics[width=1\textwidth]{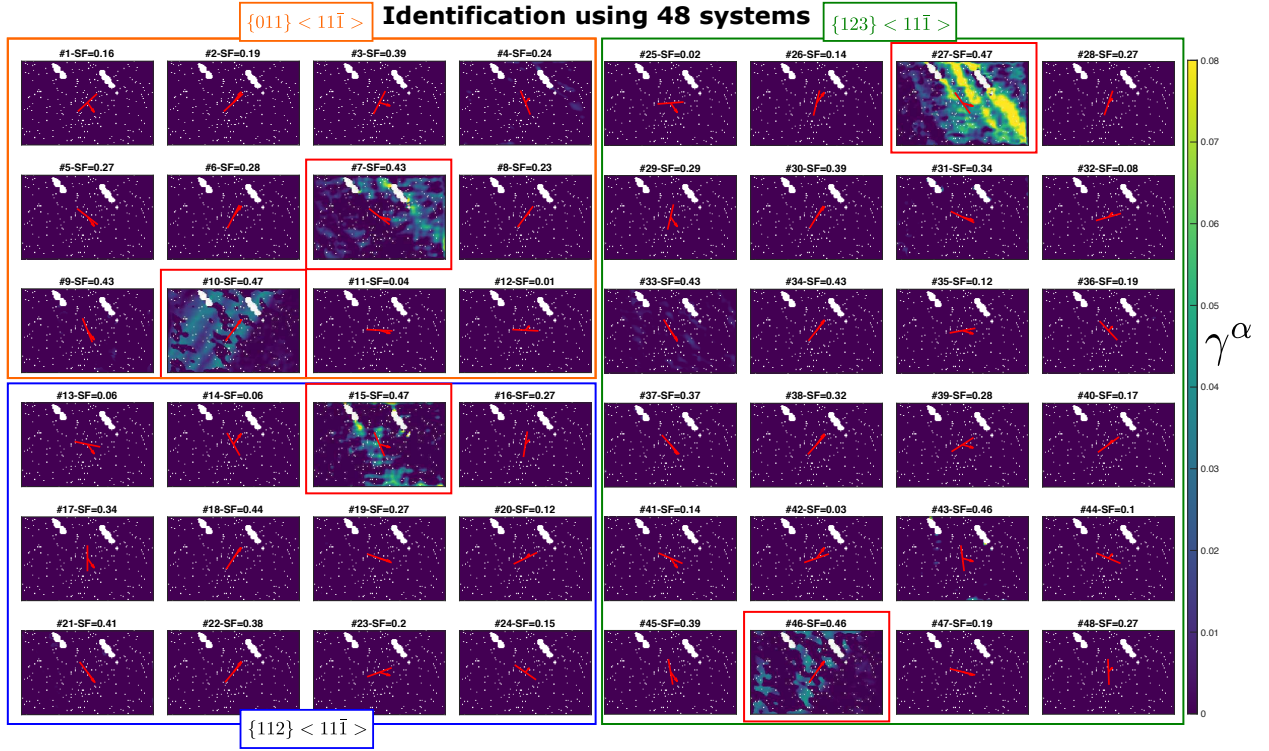}
    \caption{\small \textit{ Slip system identification results on BCC Ferrite single crystal micro-tensile test, for the zoom area in \fig \ref{fig7}d. Identified slip activity fields for all 48 ($\{011\}<11\bar{1}>$, $\{112\}<11\bar{1}>$ and $\{123\}<11\bar{1}>$) slip systems, with equal slip directions grouped in the same column, and the different boxes of different colors marking the slip families. The slip systems with significant activities are highlighted with a red square.}}
    \label{fig8}
\end{figure}

The slip system identification results, for the area marked by the red dashed rectangle in  \fig \ref{fig7}d, are displayed in \fig \ref{fig8}, showing all the 48 slip system activity fields, which are arranged with the same slip direction per column. It is clear that slip systems associated with 2 slip directions (2\ts{nd} and 3\ts{rd} column) are active, as highlighted with the red rectangles. These slip systems also have the highest SF of all considered systems. Since there is a well-defined uniaxial loading state in this specimen, the SF is a reasonable indicator for the likelihood of slip activation \cite{Tian2020}, which serves as an indirect validation of the proposed slip system identification method. Note however that the SF is not used at all in the slip identification method. 

The uncorrelated areas (due to poor DIC), shown in white in \fig\ref{fig7} and \ref{fig8}, are simply skipped in the point-by-point identification and do not introduce problems. This can be advantageous when dealing with larger deformations, in which case perhaps only part of an area is correlated successfully with DIC, on which the identification can still be applied. Moreover, assuming that the area of missing data still deforms plastically (i.e. did not damage/fracture), we could opt to interpolate these datapoints from surrounding displacement data and thereafter attempt an identification in these strongly deforming regions. However, this is outside the scope of this work.

Since there are 12 slip planes per slip direction, cross-slip is harder to identify here as there are considerably more slip planes between which it can occur than in the FCC case considered in Section \ref{sec:fcc}. Yet, for the dominant slip direction, regions do exist where there is overlap between the systems with different slip planes. Additionally, it is well known that in BCC metals slip can (appear to) occur over the Maximum Resolved Shear Stress Plane (MRSSP) through atomic scale cross-slip on two fundamental $\{011\}$ slip planes \cite{weinberger2013slip}. As such, only considering crystallographic slip planes ($\{011\}$, $\{112\}$ and $\{123\}$) may be insufficient. While the MRSSP could easily be considered in the identification as an alternative theoretical slip system, this requires knowledge on the local stress state, which is not trivially determined in most experiments.

\section{Dedicated cross-slip identification: methodology + demonstration}
\label{sec:cross-slip}
The proposed SSLIP framework is able to identify overlapping slip systems with the same slip direction but with different slip planes (and usually slip traces), which can likely be attributed to cross-slip. This effectively results in a 'combined' slip plane and trace, as was demonstrated most clearly in \fig \ref{fig5}b. However, this is likely a sub-optimal way to identify and quantify overlapping slip, and in particular cross-slip, especially when more than 2 slip traces are available for a certain slip direction, as is the case for BCC (e.g. apparent cross-slip between 3 neighbouring slip planes was observed in the BCC testcase in \fig \ref{fig8}). Therefore, we introduce here an extension of the framework which, instead of identifying the amplitudes of a range of slip systems, aims to directly identify the slip plane trace orientation (and total slip amplitude), for a given slip direction.

To this end, we construct a new point-by-point optimization problem in which the in-plane displacement gradient tensor residual L\us{2} norm is minimized:
\begin{mini!}|1|
{\gamma^1\vec{n}^{1},\gamma^2\vec{n}^{2}}{||\mathbf{H}^{exp}-(\gamma^1\vec{s}^{\:1}\otimes\vec{n}^{1} + \gamma^2\vec{s}^{\:2}\otimes\vec{n}^{2})||^{2D},  \label{eq:objfun2}}
{\label{eq:optim2}}{}
\end{mini!}
where $\mathbf{H}^{exp}$ is the measured in-plane displacement gradient tensor with its four tensor components. Whereas the 2 fixed slip directions $\vec{s}^{\: \alpha}$ are normalized, we reduce the number of variables in the optimization problem by directly identifying the vector $\gamma^\alpha\vec{n}^{\alpha}$, which is simply the in-plane normal vector multiplied with the slip amplitude.

While the identified slip plane trace normal $\vec{n}$ only contains the in-plane components of the full 3D slip plane normal, 3D knowledge of the slip direction can be employed to compute the full 3D slip plane normal, by assuming orthogonality between slip direction and slip plane normal. For cases of two overlapping bands of cross-slip, this 'dedicated cross-slip identification' method even allows for identification of the slip plane trace for two slip directions simultaneously. Both unknown slip planes $\vec{n}^1$ and $\vec{n}^2$ can be freely identified when the slip directions $\vec{s}^{\:1}$ and $\vec{s}^{\:2}$ are known (from the regular slip system identification) and are kept fixed, although these slip directions should be significantly different (in-plane).

\begin{figure}[H]
%	\centering
    \includegraphics[width=1\textwidth]{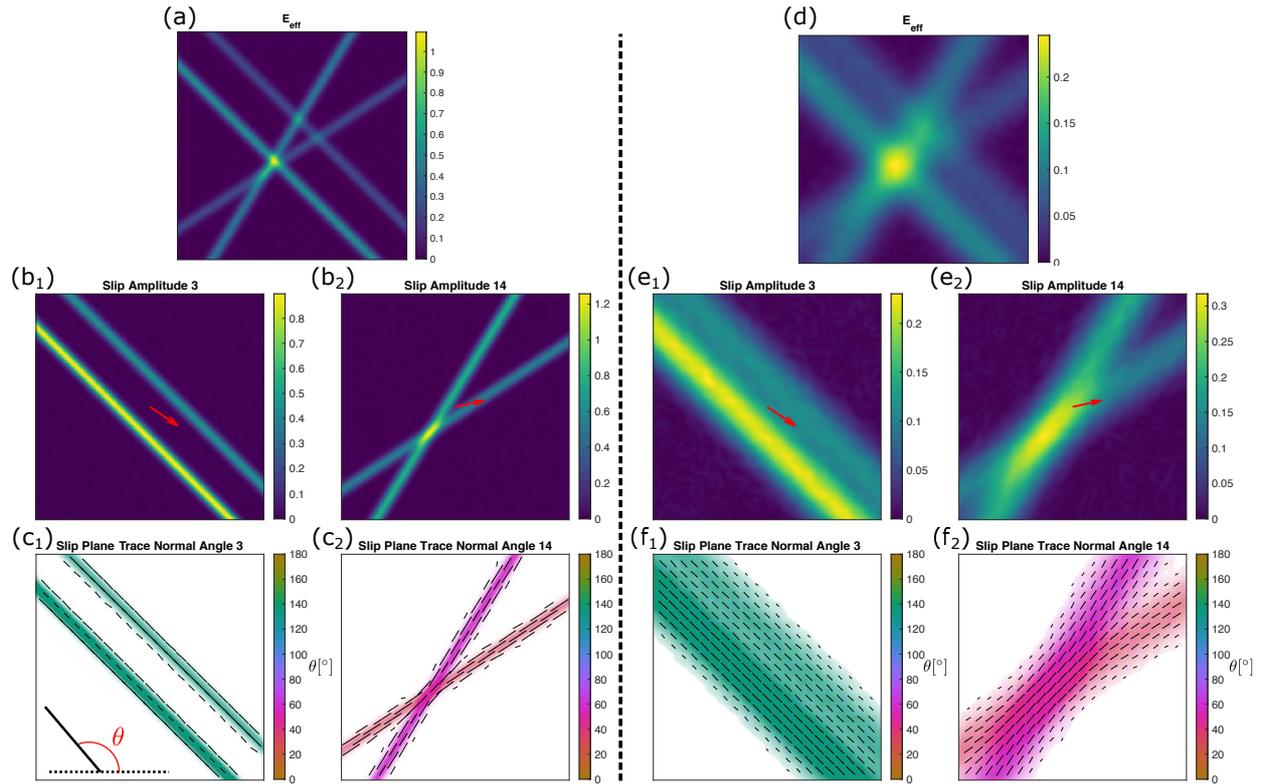}
    \caption{\small \textit{ Cross-slip identification on the HCP virtual experiments of Section \ref{sec:method} for discrete slip (a\us{1}-c\us{1}) and diffuse slip (a\us{2}-c\us{2}). (a) Equivalent strain fields, (b) slip system amplitude fields for the 2 considered slip directions, (c) slip plane trace orientation, presented as the angle between the horizontal direction and the slip plane trace, with transparency inversely scaled with the slip amplitude. We also draw (automatically, using the angle data) slip traces at intermittent locations (individual data points) to allow a more direct visual check.  
    }}
    \label{fig9}
\end{figure}

To explore the feasibility of determining a slip plane normal, and hence slip plane trace orientation, at every position, we first revisit the two HCP virtual experiments from Section \ref{sec:method}. The cross-slip identification has been performed, both for the discrete slip and the diffuse slip case, using the 2 slip directions that were identified to be active (see \fig\ref{fig3} and see the equivalent strain field in \fig\ref{fig9}a,). The identification results are shown in \fig\ref{fig9}b-c with the identified slip amplitudes ($\gamma^\alpha$) plotted in \fig\ref{fig9}b\us{1}, followed by the identified local orientation of the slip plane trace in \fig\ref{fig9}c\us{1}, presented as the angle between the horizontal direction and the slip plane trace (as drawn in the bottom-left of \fig\ref{fig9}c\us{1}). In this plot, we also draw (automatically, using the angle data) slip traces at intermittent locations to allow a more direct visual check. All the plotted traces follow the actual geometry of the slip bands. Note that these plotted traces result from individual data points in the map only. The same subfigure order is used for the diffuse virtual experiment in \fig\ref{fig9}a\us{2}-c\us{2}. For both the discrete and the diffuse case, the identified slip plane trace angles agree well with the angle of the slip bands, and thereby give confidence that the local kinematics can be used to identify the slip plane normal angle, and as such, the cross-slip plane. When curved slip would occur, due to e.g. locally varying cross-slip, our point-by-point identification method should also perform well.

\begin{figure}[H]
%	\centering
    \includegraphics[width=1\textwidth]{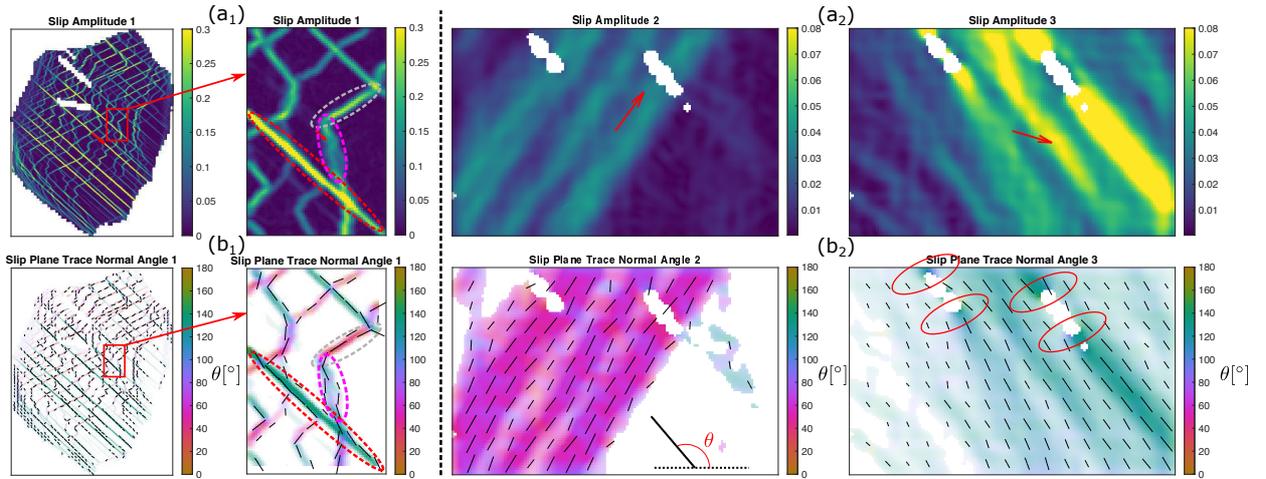}
    \caption{\small \textit{ Cross-slip identification on the FCC (a\us{1},b\us{1}) and BCC (a\us{2},b\us{2}) experimental case studies. (a) Slip system amplitude fields for one (FCC) or 2 (BCC) considered slip directions and (b) angle between horizontal direction and the slip plane orientation, with transparency inversely scaled with the slip amplitude. The red ellipses indicate areas of interesting cross-slip.
    }}
    \label{fig10}
\end{figure}

Next, we perform the cross-slip identification on Grain 1 of the FCC case study and on the BCC case study, both of which showed strong signs of cross-slip. For Grain 1 of FCC (\fig\ref{fig5}), there was only 1 slip direction identified to be active, therefore only this slip direction has been used for the identification. For the BCC case, we include both slip directions that are dominant in the slip system activity fields of \fig\ref{fig8}. \fig\ref{fig10} shows the identification of the FCC (\fig\ref{fig10}a\us{1},b\us{1}) and BCC experiment (\fig\ref{fig10}a\us{2},b\us{2}), employing the same order as in \fig \ref{fig9}. For the FCC case, we focus on the same inset as previously (\fig\ref{fig5}), in which the trace angle varies over a significant range, indicating different degrees of cross-slip. First, the red and grey ellipse again show the slip bands that are aligned with the 2 theoretical slip traces for this slip direction, showing limited local variations in the trace orientation. However, the pink ellipse indicates how the potential "cross-slip band" has a significantly different slip plane trace orientation. Moreover, local variations in trace orientation are revealed, almost between individual datapoints, indicating that the change in slip trace orientation can be captured at very high spatial resolution, not necessitating the formation of a clearly discernible slip band. 

For the BCC case study, the cross-slip identification using 2 slip directions also appears to work adequately and results in 2 overlapping slip amplitude fields (\fig\ref{fig10}a\us{2}), each with significantly different slip plane trace orientations (\fig\ref{fig10}b\us{2}), as expected according to the observed slip bands. The four red ellipses in \fig\ref{fig10}b\us{2} indicate the four tips of two small areas of missing SEM-DIC data resulting from a strong localization, which we attribute to the presence of two dominant dislocation sources. Interestingly, in these 4 ellipses, systematic variations of slip plane trace orientations can be observed, ranging from $\sim90^\circ$ to $\sim150^\circ$ (from purple blue to green/brown), while this map shows predominantly a slip plane trace angle of $\sim120^\circ$ (between blue and green). This may indicate that cross-slip may be an important mechanism near highly active dislocation sources in ferrite.

\section{SSLIP Method Limitations \& Solutions}
\label{sec:advanced}
\begin{figure}[H]
%	\centering
    \includegraphics[width=0.9\textwidth]{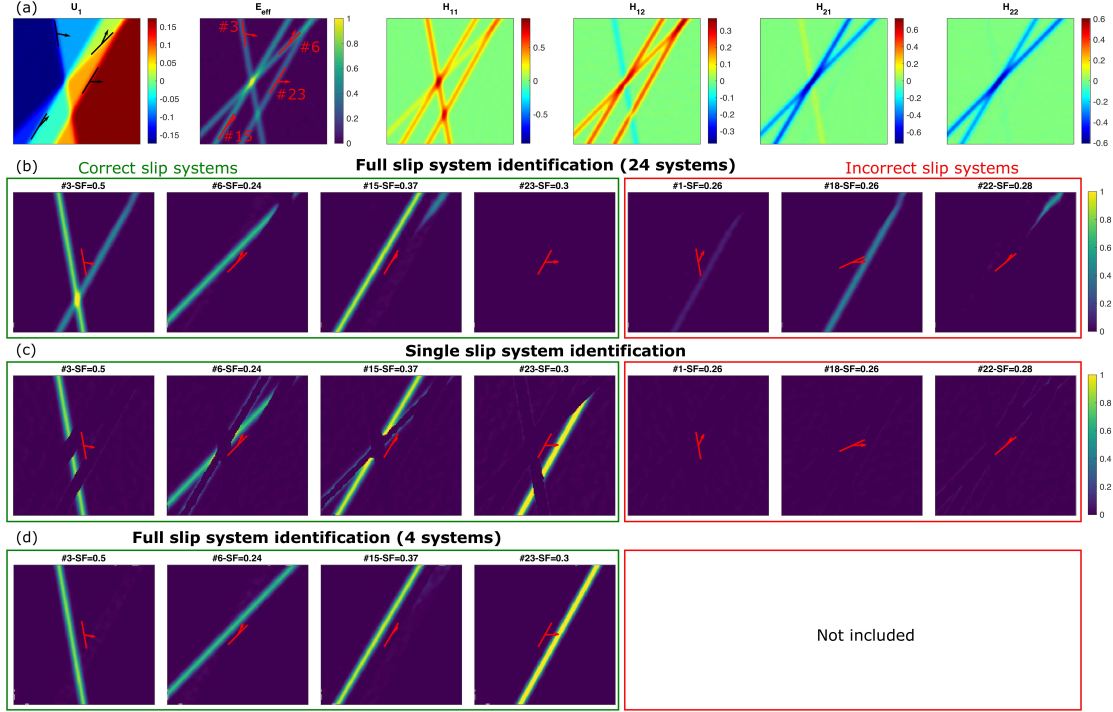}
    \caption{\small \textit{ Challenging HCP virtual experiment with 4 systems from 4 slip families, overlapping in multiple locations. (a) x-component of the displacement field, effective strain field (with the corresponding prescribed slip systems highlighted) and the 4 in-plane displacement gradient tensor components. For the full slip system identification method, we show in (b) only the slip amplitude fields where any activity occurs. The green rectangle indicates the (correct) slip systems which were used for the generation of the virtual experiment and the red rectangle shows incorrectly identified slip amplitude fields. (c) Single slip system identification results. (d) Slip system identification results in which only the 4 active systems, as identified through single slip identification, were included.
    }}
    \label{fig11}
\end{figure}

For both the virtual and experimental case studies investigated up to this point, there were no inconsistencies in the form of wrongly identified slip activities. For all test cases treated above, more than 4 potential slip systems were included. The optimization problem allowed for the solution if the sum of the slip amplitudes was minimized, which appeared to work well for the considered case studies up to this point. However, HCP can be too complex in some situations involving 24 slip systems, in the form of 3 Basal, 3 Prismatic, 12 1\ts{st} and 6 2\ts{nd} order $<c+a>$ Pyramidal slip systems. These slip systems consist of a total of 9 unique slip directions and 15 unique slip planes. \fig\ref{fig11}a shows a dedicated HCP virtual experiment with a different crystal orientation and more active slip systems, as compared to the virtual experiment in Section \ref{sec:method}. For this new virtual experiment, we generate a challenging case of 4 different active slip systems from 4 different slip families: 1 Basal ($\#3$), 1 Prismatic ($\#6$), 1 1\ts{st} order ($\#15$) and 1 2\ts{nd} order ($\#23$) Pyramidal slip system. Moreover, there is also considerable overlap between 3 slip systems and the displacement gradient tensor component fields of systems $\#6$ and $\#15$ are almost equal, except for the $H_{11}$ component. \fig\ref{fig11}b shows the results of the slip system identification including all 24 slips systems, plotting only the amplitude fields of slip systems that reveal any activity. The activity of slip systems $\#3$, $\#6$ and $\#15$ are correctly captured. However, the activity of slip system $\#23$ is not recovered in the amplitude field of $\#23$, but instead it is identified as a combination of systems $\#1$, $\#3$ and $\#18$. This can be attributed to the high number of in-plane linearly dependent slip system kinematics. In such complex cases, one could also verify the identified amplitude field through comparison between the visible slip band and the theoretical slip trace angle, which clearly is not correct in several of the amplitude fields ($\#1$, $\#3$ and $\#18$).

For such cases with a high number of in-plane linearly dependent slip systems, we postulate that the complexity of the optimization problem needs to be reduced, by limiting the number of slip systems that are considered. This will ensure that the minimization of the sum of slip amplitudes is less dominant and that the constraint of the displacement gradient tensor residual predominantly drives the solution. Multiple strategies can be employed to perform a preselection, such as (i) selecting a small number of systems after the full identification, (ii) considering only the combination of slip systems in, e.g., 1 or 2 families at the same time (especially for HCP, e.g., based on prior knowledge), (iii) minimize the dissipated energy, instead of the slip amplitudes (as suggested in Section \ref{sec:method}) or (iv) aim to identify if there are locations in the deformation field that can be adequately captured by the kinematics of a single slip system (highlighting the importance of that system), which we will explore here. For this "single slip system identification", we solve the "regular" optimization problem, as described in Section \ref{sec:method}, for every slip system $\alpha$ separately. Since this is an overdetermined problem, the residual L\us{2} norm of the optimization problem ($||\mathbf{H}^{exp}-\mathbf{H}^{theor}||^{2D}$) is not guaranteed to reach the threshold $H_{thresh}$, since the slip amplitude for all slip systems except one is set at 0. Note that the minimization of the sum of amplitudes reduces to the minimization of a single amplitude, which effectively means that the optimization problem is almost completely driven by the constraints. Also note that this approach cannot give a viable solution on positions of overlapping slip, while extra care needs to be taken not to lower the threshold below the SEM-DIC noise level. The result of the single slip system identification is shown in \fig\ref{fig11}c, showing a large improvement, as the 4 correct slip systems are indeed identified in regions without overlap, while the remaining (incorrect) slip systems show no real slip activity. However, as predicted, the slip activity is completely missing in the overlap (intersecting) regions, while the presence of artificial noise also slightly degrades the identification quality. Nevertheless, even with these limitations, it is clear that the single slip system identification does provide strong indications on the activation of individual systems, to be exploited for further analysis.

Once the 4 active slip systems have been identified using single slip system identification, only these 4 slip systems (in this case $\#3$, $\#6$, $\#15$ and $\#23$) are included in the "main" slip system identification method, for which the results are shown in \fig\ref{fig11}d, which agrees well with the generated slip steps in the virtual experiment. On balance, we recommend caution when performing slip identification on HCP, especially when considering multiple slip families. Single slip system identification, or even multiple slip identification attempts with pairs of slip systems, is suggested as an initial step in the analysis, to be sure.

\section{Conclusions}
\label{sec:conclusions}
Slip system identification methods are an important tool for the analysis of the micromechanics of polycrystalline metals, but their applicability was often limited to discrete slip bands, for which the slip traces could clearly be identified from the strain fields. More complex manifestations of crystallographic slip, such as cross-slip, curved slip, diffuse slip and/or overlapping slip, often lack a clearly identifiable slip trace direction also due to the limited DIC resolution and therefore require a novel and local identification approach that does not rely on the identification of slip traces (and does not use the Schmid Factors of the slip systems). 

To this end, this paper proposed a \textbf{S}lip \textbf{S}ystems based \textbf{I}dentification of \textbf{L}ocal \textbf{P}lasticity (SSLIP) framework, in which (SEM-)DIC displacement fields are used to perform a local point-by-point identification. The underlying assumption is that the in-plane displacement gradient tensor components, computed directly from the displacement field, contain the in-plane deformation kinematics that conforms to a combination of active slip system kinematics. Subsequently, an optimization problem is solved on each point of the SEM-DIC map which results in the amplitude of all considered slip systems, i.e. the slip activity fields. Even DIC maps with missing data, e.g. due to large strains, do not pose any problems.

Several case studies were investigated by which the validity and applicability of the SSLIP method was clearly demonstrated:
\begin{itemize}
\item a virtual HCP case study (with added experimental noise) of discrete and diffuse overlapping slip bands, over 3 slip systems, was used to explain, demonstrate and validate the methodology, showing that the method performs well when using up to 24 slip systems simultaneously for identification;
\item an experimental FCC case study of a Ni-based superalloy (taken from literature \cite{harte2020statistical,harte2020effect}), in which discrete slip, cross-slip, diffuse slip and overlapping slip are all present, showed that the local nature of the slip system identification method is paramount for a correct identification of these challenging slip mechanisms, while a few isolated discrete slip bands allowed an indirect validation of the method;
\item an SEM-DIC micro-tensile test on a BCC ferrite single crystal showed diffuse slip that could be correctly identified using 48 slip systems simultaneously.
\end{itemize}

Cross-slip was indirectly revealed using the slip system identification method, by assessing overlap between slip systems with the same slip direction. Moreover, an extension to the framework, dedicated for the identification of potential cross-slip (where dislocations can glide over any irrational plane), was also proposed and validated. To this end, an alternative optimization scheme was constructed in which the slip plane trace orientation is directly identified, when one or two cross-slip directions are known (from the slip system identification method) and kept fixed. The cross-slip identification framework was validated on the same virtual HCP experiment, correctly identifying the slip plane trace orientation at each position. Furthermore, application to the FCC and BCC experiments yielded detailed insights into the potential cross-slip behaviour.

For assessment of the limitations of the proposed slip system identification framework, we explored a highly challenging virtual case with 4 different and overlapping slip systems, for which the slip system identification method only found a partially correct solution, due to the underdetermined nature of the optimization problem and the complexity of HCP plasticity from 4 slip families. A solution was provided in the form of a two-step approach. First a single slip system identification was performed in order to confidently identify areas where single slip systems are dominant. Next, the identified single slip systems were solely used to perform a full combined identification, resulting in the expected solution.

Finally, the identification could be extended to include additional measurement data to improve robustness and accuracy. For instance, (quasi-)3D DIC, also know as Digital Height Correlation \cite{kleinendorst2016adaptive}, could be employed on nanoscale Atomic Force Microscopy (AFM) measurements, to access additional 3D displacement gradient tensor components, which can directly be included in the minimization function. Additionally, simultaneous \textit{in-situ} EBSD (along DIC) \cite{edwards2022mapping} can yield local lattice rotations, and in particular their rotation axes, which might also contribute to the identification.

\section*{Author Contributions (CRediT)}
\textbf{Tijmen Vermeij:} Conceptualization, Methodology, Software, Validation, Investigation, Writing - Original Draft, Visualization

\textbf{Ron Peerlings:} Methodology, Writing - Review \& Editing, Supervision, Funding Acquisition

\textbf{Marc Geers:} Methodology, Resources, Writing - Review \& Editing, Supervision, Funding Acquisition

\textbf{Johan Hoefnagels:} Methodology, Validation, Resources, Writing - Review \& Editing, Supervision, Funding Acquisition

\section*{Acknowledgements}
The authors thank the anonymous reviewers for their constructive comments. We acknowledge Tim J.J. Ramirez y Cantador and Jorn A.C. Verstijnen for discussions and experimental support. João Quinta da Fonseca (and colleagues) is acknowledged for providing the open-access data of the Ni superalloy. 

This research was carried out as part of the "UNFAIL" project, under project number S17012b in the framework of the Partnership Program of the Materials innovation institute M2i (www.m2i.nl) and the Netherlands Organization for Scientific Research (http://www.nwo.nl).

\section*{Code and Data availability}
The Matlab code for the full SSLIP method, with several examples, is available on Github:\\
 \url{https://www.github.com/TijmenVermeij/SSLIP}. The full datasets are available upon request.

\centering
\noindent\rule{8cm}{0.4pt}

\clearpage

%\bibliography{lib}

\end{document}